\documentclass{aa}
\usepackage{graphicx}
\usepackage{natbib}
\usepackage{txfonts}

\begin{document}

\title{On the nature of GRB 050509b: a disguised short GRB}

\author{
G. De Barros\inst{1,2}
\and
L. Amati\inst{2,3}
\and
M.G. Bernardini\inst{4,1,2}
\and
C.L. Bianco\inst{1,2}
\and
L. Caito\inst{1,2}
\and
L. Izzo\inst{1,2}
\and
B. Patricelli\inst{1,2}
\and
R. Ruffini\inst{1,2,5}
}

\authorrunning{G. de Barros et al.}

\titlerunning{On the nature of GRB 050509b}

\institute{
Dipartimento di Fisica and ICRA, Universit\`a di Roma ``La Sapienza'', Piazzale Aldo Moro 5, I-00185 Roma, Italy. E-mails: [maria.bernardini;bianco;letizia.caito;luca.izzo;ruffini]@icra.it
\and
ICRANet, Piazzale della Repubblica 10, I-65122 Pescara, Italy. E-mails: [gustavo.debarros;barbara.patricelli]@icranet.org
\and
Italian National Institute for Astrophysics (INAF) - IASF Bologna, via P. Gobetti 101, 40129 Bologna, Italy. E-mail: amati@iasfbo.inaf.it
\and
Italian National Institute for Astrophysics (INAF) - Osservatorio Astronomico di Brera, via Emilio Bianchi 46, I-23807 Merate (LC), Italy.
\and
ICRANet, Universit\'e de Nice Sophia Antipolis, Grand Ch\^ateau, BP 2135, 28, avenue de Valrose, 06103 NICE CEDEX 2, France.
}

\date{}

\abstract
{
GRB 050509b, detected by the \emph{Swift} satellite, is the first case where an X-ray afterglow has been observed associated with a short gamma-ray burst (GRB). Within the fireshell model, the canonical GRB light curve presents two different components: the proper-GRB (P-GRB) and the extended afterglow. Their relative intensity is a function of the fireshell baryon loading parameter $B$ and of the CircumBurst Medium (CBM) density ($n_{CBM}$). In particular, the traditionally called short GRBs can be either ``genuine'' short GRBs (with $B \lesssim 10^{-5}$, where the P-GRB is energetically predominant) or ``disguised'' short GRBs (with $B \gtrsim 3.0 \times 10^{-4}$ and $n_{CBM}\ll1$, where the extended afterglow is energetically predominant).
}
{
We verify whether GRB 050509b can be classified as a ``genuine'' short or a ``disguised'' short GRB, in the fireshell model.
}
{
We investigate two alternative scenarios. In the first, we start from the assumption that this GRB is a ``genuine'' short burst. In the second attempt, we assume that this GRB is a ``disguised'' burst.
}
{
If GRB 050509b were a genuine short GRB, there should initially be very hard emission which is ruled out by the observations. The analysis that assumes that this is a disguised short GRB is compatible with the observations. The theoretical model predicts a value of the extended afterglow energy peak that is consistent with the Amati relation.
}
{
GRB 050509b cannot be classified as a ``genuine'' short GRB. The observational data are consistent with a ``disguised'' short GRB classification, i.e., a long burst with a weak extended afterglow ``deflated'' by the low density of the CBM. We expect that all short GRBs with measured redshifts are disguised short GRBs because of a selection effect: if there is enough energy in the afterglow to measure the redshift, then the proper GRB must be less energetic than the afterglow. The Amati relation is found to be fulfilled only by the extended afterglow excluding the P-GRB.
}

\maketitle

\section{Introduction}

The traditional classification of gamma ray bursts (GRBs) is based on the observed time duration of the prompt emission measured with the criterion of ``$T_{90}$'', which is the time duration in which the cumulative counts increase from $5\%$ to $95\%$ above the background, encompassing $90\%$ of the total GRB counts. This parameter shows that there are two groups of GRBs, the short ones with $T_{90}<2$ s, and the long ones with $T_{90}>2$ s. This analysis motivated the standard classification in the literature of short and long GRBs \citep{1992grbo.book..161K,1992AIPC..265..304D,1993ApJ...413L.101K}.

\begin{figure}
\centering
\includegraphics[width=\hsize]{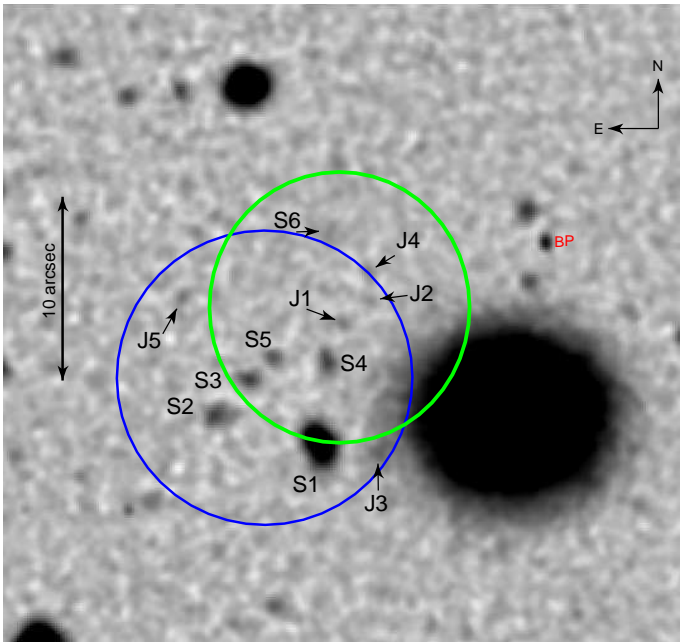}
\caption{Keck LRIS G-band image, zoomed to show the XRT error circle. The larger, blue circle is the revised XRT position from \citet{2005GCN..3395....1R}; the smaller, green circle to the west and north of that is the $2\sigma$ confidence region of the XRT position computed in \citet{2006ApJ...638..354B}. The 11 sources consistent with the \citet{2005GCN..3395....1R} X-ray afterglow localization are labeled in the image. North is up and east is to the left. G1 is  the large galaxy to the west and south of the XRT. Bad pixel locations are denoted with ``BP''. Figure reproduced from \citet{2006ApJ...638..354B} with the kind permission of J. Bloom and of the AAS.}
\label{fbloom}
\end{figure}

The observations of GRB 050509b by BAT and XRT on board the \emph{Swift} satellite \citep[see][]{2004ApJ...611.1005G,2005SSRv..120..165B} represent a new challenge to the classification of GRBs as long and short, since it is the first short GRB associated with an afterglow \citep{2005Natur.437..851G}. Its prompt emission observed by BAT lasts 40 milliseconds, but it also has an afterglow in the X-ray band observed by XRT, which begins 100 seconds after the BAT trigger (time needed to point XRT to the position of the burst) and lasts until $\approx$ 1000 seconds. It is located 40 kpc away from the center of its host galaxy \citep[][, see Fig. \ref{fbloom}]{2006ApJ...638..354B}, which is a luminous, non-star-forming elliptical galaxy with redshift $z=0.225$ \citep{2005Natur.437..851G}. Although an extensive observational campaign has been performed using many different instruments, no convincing optical-IR candidate afterglow nor any trace of any supernova has been found associated with GRB 050509b \citep[see][]{2005GCN..3401....1C,2005GCN..3521....1B,2005ApJ...630L.117H,2005A&A...439L..15C,2005GCN..3386....1B,2005GCN..3417....1B,2006ApJ...638..354B}. An upper limit in the $R$-band $18.5$ days after the event onset imply that the peak flux of any underlying supernova should have been $\sim 3$ mag fainter than the one observed for the type Ib/c supernova SN 1998bw associated with GRB 980425, and $2.3$ mag fainter than a typical type Ia supernova (\citealp{2005A&A...439L..15C}, see also \citealp{2005ApJ...630L.117H}). An upper limit to the brightening caused by a supernova or supernova-like emission has also been established at $8.17$ days after the GRB: $R_c \sim 25.0$ mag \citep{2006ApJ...638..354B}. While some core-collapse supernovae might be as faint as (or fainter than) this limit \citep{2007Natur.449E...1P}, the presence of this supernova in the outskirts of an elliptical galaxy would be truly extraordinary \citep{2005A&A...433..807M,2005PASP..117..773V}.

Unfortunately, we cannot obtain exhaustive observational constraints for this GRB because XRT data are missing in-between the first 40 milliseconds and 100 seconds. However, this makes the theoretical work particularly interesting, because we can infer from first principles some characteristics of the missing data, which are inferred by our model, and consequently reach a definite understanding of the source. This is indeed the case, specifically, for the verification of the Amati relation \citep{2002A&A...390...81A,2006MNRAS.372..233A,2009A&A...508..173A} for these sources as we see in section \ref{amati}.

GRB 050509b is an example that the usual classification is at least incomplete. Within the fireshell model, we propose three classes of GRBs: long, genuine short and disguised short \citep[][and references therein]{2009AIPC.1132..199R}. We have a well-defined way of differentiating between the classes, which is based on two parameters, the baryon loading parameter $B$ and the CircumBurst Medium (CBM) number density $n_{CBM}$ (see next section), that help to make the classification clearer. In this paper, we analyze GRB 050509b within the fireshell model. We proceed with the identification of the two basic parameters, $B$ and $n_{CBM}$, within two different scenarios. We first investigate the ``ansatz'' that this GRB is the first example of a ``genuine'' short bursts. After disproving this possibility, we show that this GRB is indeed another example of a disguised short burst.

In the next section, we briefly introduce the fireshell model and explain the classification, in section \ref{analysis} we show the analysis of the data, in section \ref{amati} we present the theoretical spectrum and the study of the fulfillment of the Amati relation, in section \ref{sec:discussions} we comment on the results, and in section \ref{sec:conclusions} we finally present our conclusions.

\section{The fireshell model}\label{fireshell}

Within the fireshell model \citep{2002ApJ...581L..19R,2004IJMPD..13..843R,2005IJMPD..14...97R,2009AIPC.1132..199R,2005ApJ...620L..23B,2005ApJ...633L..13B}, all GRBs originate from an optically thick $e^\pm$ plasma of total energy $E_{tot}^{e^\pm}$ in the range $10^{49}$--$10^{54}$ ergs and a temperature $T$ in the range $1$--$4$ MeV. After an early expansion, the $e^\pm$-photon plasma reaches thermal equilibrium with the engulfed baryonic matter $M_B$ described by the dimensionless parameter $B=M_{B}c^{2}/E_{tot}^{e^\pm}$, which must be $B < 10^{-2}$ to allow the fireshell to expand further. As the optically thick fireshell composed of $e^\pm$-photon-baryon plasma self-accelerates to ultrarelativistic velocities, it finally reaches the transparency condition. A flash of radiation is then emitted. This represents the proper-GRB (P-GRB). The amount of energy radiated in the P-GRB is only a fraction of the initial energy $E_{tot}^{e^\pm}$. The remaining energy is stored in the kinetic energy of the optically thin baryonic and leptonic matter fireshell that, by inelastic collisions with the CBM, gives rise to multiwavelength emission. This is the extended afterglow.

\begin{figure}
\centering
\includegraphics[width=\hsize]{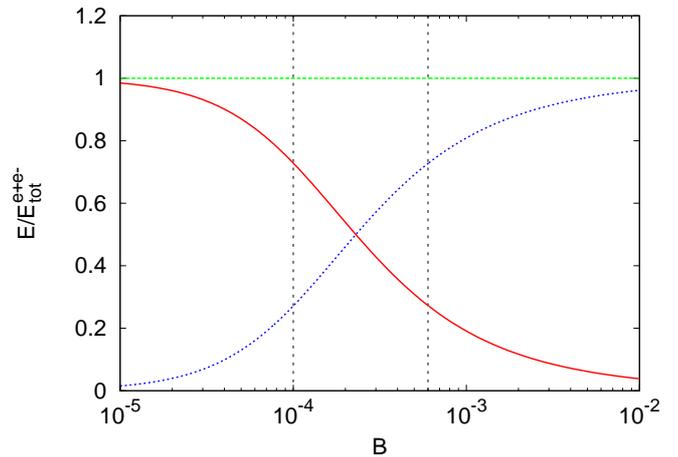}
\caption{The dashed (blue) curve is the energy emitted in the extended afterglow, the solid (red) curve is the energy emitted in the P-GRB, their sum is $E_{tot}^{e^\pm}$. From left to right, the first vertical line corresponds to the value of $B=1.0\times 10^{-4}$ of scenario 1, the second to the value of $B=6.0\times 10^{-4}$ of scenario 2 (see Sec. \ref{analysis}).}\label{B}
\end{figure}

Within this model, the value of $B$ strongly affects the ratio of the energetics of the P-GRB to the kinetic energy of the baryonic and leptonic matter within the extended afterglow phase. It also affects the time separation between the corresponding peaks \citep{2009AIPC.1132..199R}. For baryon loading $B \lesssim 10^{-5}$, the P-GRB component is always energetically dominant over the extended afterglow (see Fig. \ref{B}). In the limit $B \rightarrow 0$, it gives rise to a ``genuine'' short GRB. Otherwise, when $3.0\times 10^{-4} \lesssim B \leq 10^{-2}$, the kinetic energy of the baryonic and leptonic matter, and consequently the extended afterglow emission, predominates with respect to the P-GRB \citep{2002ApJ...581L..19R,2007A&A...474L..13B}. Since the ``critical'' value of $B$ corresponding to the crossing point in Fig. \ref{B} is a slowly varying function of the total energy $E_{tot}^{e^\pm}$, for $10^{-5} \lesssim B \lesssim 3.0 \times 10^{-4}$ the ratio of the total energies of the P-GRB and the extended afterglow is also a function of $E_{tot}^{e^\pm}$.

The extended afterglow luminosity in the different energy bands is governed by two quantities associated with the environment: the CBM density profile, $n_{CBM}$, and the ratio of the effective emitting area $A_{eff}$ to the total area $A_{tot}$ of the expanding baryonic shell, ${\cal R}= A_{eff}/A_{tot}$. This second parameter takes into account the CBM filamentary structure \citep{2004IJMPD..13..843R}. Typical values of ${\cal R}$ ranges between $10^{-10}$ and $10^{-6}$ \citep[see e.g.][]{2007A&A...474L..13B,2005ApJ...634L..29B,2009A&A...498..501C,2010A&A...521A..80C,2007A&A...471L..29D,2006ApJ...645L.109R}.

The emission from the baryonic matter shell is spherically symmetric. This allows us to assume, to a first approximation, a modeling of thin spherical shells for the CBM distribution and consequently only consider their radial dependence \citep{2002ApJ...581L..19R}. The emission process is assumed to be thermal in the comoving frame of the shell \citep{2004IJMPD..13..843R}. The observed GRB non-thermal spectral shape is produced by the convolution of a very large number of thermal spectra with different temperatures and different Lorentz and Doppler factors. This convolution is performed over the surfaces of constant arrival times for the photons at the detector \citep[EQuiTemporal Surfaces, EQTSs;][]{2005ApJ...620L..23B,2005ApJ...633L..13B} encompassing the total observation time. The fireshell model does not address the plateau phase described by \citet{2006ApJ...642..389N}, which may not be related to the interaction of the single baryonic shell with the CBM \citep{ba10}.

In the context of the fireshell model, we considered a new class of GRBs, pioneered by \citet{2006ApJ...643..266N}. This class is characterized by an occasional softer extended emission after an initial spike-like emission. The softer extended emission has a peak luminosity lower than the one of the initial spike-like emission. As shown in the prototypical case of GRB 970228 \citep{2007A&A...474L..13B} and then in both GRB 060614 \citep{2009A&A...498..501C} and GRB 071227 \citep{2010A&A...521A..80C}, we can identify the initial spike-like emission with the P-GRB and the softer extended emission with the peak of the extended afterglow. A crucial point is that the time-integrated extended afterglow luminosity (i.e. its total radiated energy) is much higher than the P-GRB one. This unquestionably identifies GRB 970228 and GRB 060614 as canonical GRBs with $B > 10^{-4}$. The consistent application of the fireshell model allows us to infer the CBM filamentary structure and average density, which, in that specific case, is $n_{CBM} \sim 10^{-3}$ particles/cm$^3$, typical of a galactic halo environment \citep{2007A&A...474L..13B}. This low CBM density value explains the peculiarity of the low extended afterglow peak luminosity and its more protracted time evolution. These features are not intrinsic to the progenitor, but depend uniquely on the peculiarly low value of the CBM density. This led us to expand the traditional classification of GRBs to three classes: ``genuine'' short GRBs, ``fake'' or ``disguised'' short GRBs, and the remaining ``long duration'' ones.

A CBM density $n_{CBM} \sim 10^{-3}$ particles/cm$^3$ is typical of a galactic halo environment, and GRB 970228 was indeed found to be in the halo of its host galaxy \citep{1997Natur.387R.476S,1997Natur.386..686V}. We therefore proposed that the progenitors of this new class of disguised short GRBs are merging binary systems, formed by neutron stars and/or white dwarfs in all possible combinations, which spiraled out from their birth place into the halo \cite[see][]{2007A&A...474L..13B,2009A&A...498..501C,KMG11}. This hypothesis can also be supported by other observations. Assuming that the soft-tail peak luminosity is directly related to the CBM density, short GRBs displaying a prolonged soft tail should have a systematically smaller offset from the center of their host galaxy. Some observational evidence was found in this sense \citep{2008MNRAS.385L..10T}. However, the present sample of observations does not enable us to derive any firm conclusion that short GRBs with extended emission have smaller physical offsets than those without extended emission \citep{2010ApJ...708....9F,2011NewAR..55....1B}.

\section{Data analysis of GRB 050509b}\label{analysis}

\subsection{Scenario 1}\label{anal1}

We first attempt to analyze GRB 050509b under the scenario that assumes it is a ``genuine'' short GRB, namely a GRB in which more than $50\%$ of the total energy is emitted in the P-GRB. This would be the first example of an identified ``genuine'' short GRB.

Within our model, the only consistent solution that does not contradict this assumption leads to the interpretation that all the data belongs to the extended afterglow phase; the BAT data of the prompt emission \citep[see figure 2 in][]{2005Natur.437..851G} are then the peak of the extended afterglow, and the XRT data represents the decaying phase of the extended afterglow (which in the literature is simply called ``the afterglow'', see section \ref{fireshell}). 

\begin{figure}
\centering
\includegraphics[width=\hsize]{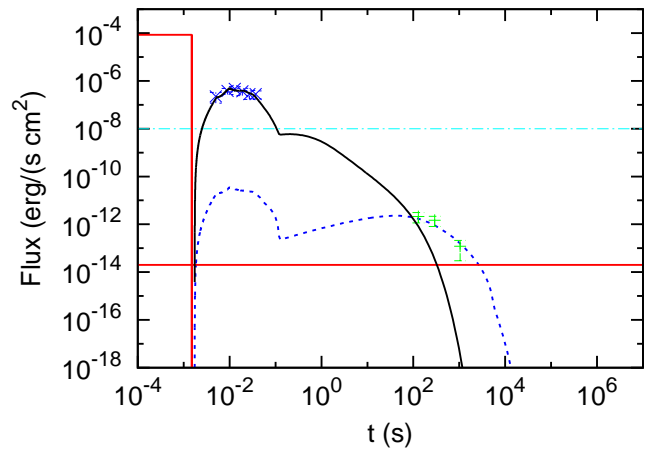}
\caption{Our numerical simulation within scenario 1, assuming that GRB 050509b is a ``genuine'' short GRB, i.e. that the P-GRB is energetically predominant over the extended afterglow. The BAT data (crosses) are interpreted as the peak of the extended afterglow. In this case, the predicted P-GRB (solid rectangle) total energy is more than twice the extended afterglow one. The solid line is the theoretical light curve in the 15-150 keV energy band, and the dashed one is the theoretical light curve in the 0.3-10 keV energy band. The dot-dashed horizontal line represents the BAT threshold and the solid horizontal one represents the XRT threshold.}
\label{fig:After}
\end{figure}

In figure \ref{fig:After}, we show the result of this analysis. We obtained the following set of parameters: $E_{tot}^{e^\pm}=2.8 \times 10^{49}$ erg, $B=1.0 \times 10^{-4}$, and $n_{CBM}=1.0 \times 10^{-3}$ particles/cm$^3$. These parameters would imply, however, that the energy emitted in the P-GRB should be almost $72\%$ of the total value. This P-GRB should have been clearly observable, and has not been detected. Consequently, this scenario is ruled out and we conclude that GRB 050509b cannot be interpreted as a ``genuine'' short GRB.

\subsection{Scenario 2}\label{anal2}

\begin{figure}
\centering
\includegraphics[width=\hsize]{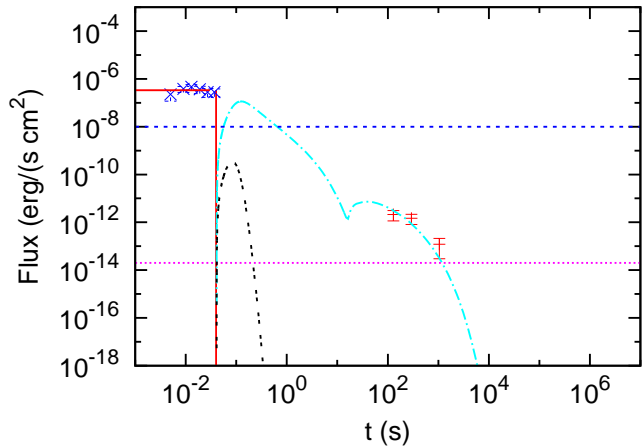}
\caption{Our numerical simulation within scenario 2, assuming that the extended afterglow is energetically predominant over the P-GRB. In this case, the predicted P-GRB (solid rectangle) is less then twice the extended afterglow. We interpret the BAT data (crosses) as the P-GRB and the XRT data as the extended afterglow. The P-GRB has just $28\%$ of the total energy. The double-dashed line is the theoretical light curve in the band $15$--$150$ keV, and the dot-dashed line is the theoretical light curve in the band $0.3$--$10$ keV. The two horizontal lines are from above to below: the BAT threshold and the XRT threshold.}
\label{fig:P-GRB}
\end{figure}

We now analyze GRB 050509b under the alternative scenario that assumes the energy of the extended afterglow is higher than the P-GRB one. Within our model, the only consistent solution that does not contradict this assumption leads to the prompt emission observed by BAT \citep[see Fig. 2 in][]{2005Natur.437..851G} being interpreted as the P-GRB, and the X-ray decaying afterglow data observed by XRT being interpreted as the extended afterglow.

In Fig. \ref{fig:P-GRB}, we show the result of this analysis. We obtained the parameters $E_{tot}^{e^\pm}=5.52 \times 10^{48}$ erg, $B=6 \times 10^{-4}$, and an almost constant CBM density $n_{CBM}=1.0 \times 10^{-3}$ particles/cm$^3$. The low value of the number density is justified by the GRB being located $40$ kpc away from the center of the host galaxy \citep[][, see Fig. \ref{fbloom}]{2006ApJ...638..354B}. The initial value of ${\cal R}$ is quite large, ${\cal R} = 1.2\times 10^{-1}$, which indicates that there is a very homogeneous CBM in the region close to the progenitor system. However, at $t\approx 10$ seconds, corresponding to a fireshell radius of $\sim 2\times 10^{16}$ cm, the effective area of interaction between the expanding plasma and the CBM drops six orders of magnitude and we have ${\cal R} = 3 \times 10^{-7}$, a value pointing to the typical CBM filamentary structure also encountered in other sources (see Sec. \ref{fireshell}). The P-GRB has an estimated energy of $E_{P-GRB}=28\%E_{tot}^{e^\pm}$, which means that $72\%$ of the energy is released in the extended afterglow. The peak of the extended afterglow, theoretically predicted by our model in figure \ref{fig:P-GRB}, was not observed by BAT, since the energy was below its threshold, and also not observed by XRT, since unfortunately its data collection started only 100 seconds after the BAT trigger.

Following our classification, therefore, due to the values of the baryon loading and of the CBM density, as well as due to the offset with respect to the host galaxy, GRB 050509b is consistent with being another example of a disguised short GRB. This follows the previous identification of GRB 970228 \citep{2007A&A...474L..13B} GRB 060614 \citep{2009A&A...498..501C} and GRB 071227 \citep{2010A&A...521A..80C}.

\section{The theoretical spectrum and Amati relation}\label{amati}

We turn now to the most interesting aspects of our theoretical work, namely the possibility of inferring some characteristics of the missing data and finally the nature of the burst from first principles.
The most effective tool for determining the nature and, then, interpreting the different classes of GRBs, is the Amati relation \citep{2002A&A...390...81A,2006MNRAS.372..233A,2009A&A...508..173A}. This empirical spectrum-energy correlation states that the isotropic-equivalent radiated energy of the prompt emission $E_{iso}$ is correlated with the cosmological rest-frame $\nu F_{\nu}$ spectrum peak energy $E_{p,i}$: $E_{p,i}\propto (E_{iso})^{a}$, where $a \approx 0.5$ and a dispersion $\sigma(\log_{Ep}) \sim 0.2$. The Amati relation holds only for long duration bursts, while short ones, as it has been possible to prove after the ``afterglow revolution'' and the measurement of their redshift, are inconsistent with it \citep{2006MNRAS.372..233A,2009A&A...508..173A}.

This dichotomy can naturally be explained by the fireshell model. As recalled in Sect. \ref{fireshell}, within this theoretical framework the prompt emission of long GRBs is dominated by the peak of the extended afterglow, while that of the short GRBs is dominated by the P-GRB. Only the extended afterglow emission follows the Amati relation \citep[see][]{2008A&A...487L..37G,2010A&A...521A..80C}. Therefore, all GRBs in which the P-GRB provides a negligible contribution to the prompt emission (namely the long ones, where the P-GRB is at most a small precursor) fulfill the Amati relation, while all GRBs in which the extended afterglow provides a negligible contribution to the prompt emission (namely the short ones) do not \citep[see][]{2007A&A...474L..13B,2008AIPC..966....7B,2008A&A...487L..37G,2009A&A...498..501C,2010A&A...521A..80C}. As a consequence, for disguised short bursts the two components of the prompt emission must be analyzed separately. The first spikelike emission alone, which is identified with the P-GRB, should not follow the Amati relation; the prolonged soft tail, which is identified with the peak of the extended afterglow, should instead follow the Amati relation. This has been confirmed in the cases of GRB 060614 and GRB 071227 \citep{2010A&A...521A..80C}.

\begin{figure}
\centering
\includegraphics[width=\hsize]{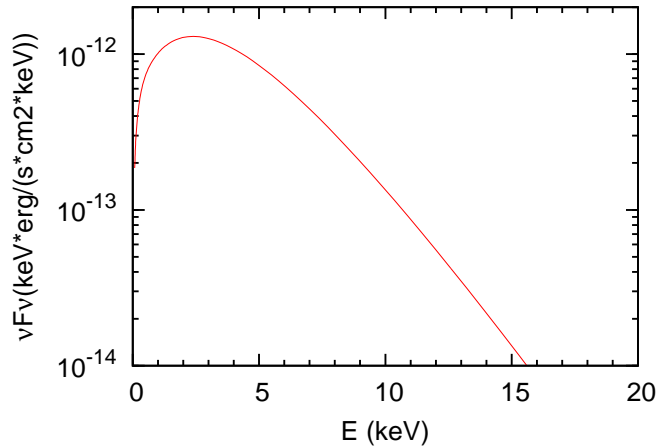}
\caption{Our theoretical spectrum in the observer frame integrated over the entire extended afterglow up to $10^4$ s, see also \citet{2008A&A...487L..37G}.}
\label{fig:spectrum}
\end{figure}
     
Owing to the lack of extended afterglow observational data before $100$ seconds, there is no way to confirm whether this source follows the Amati relation on an observational ground. To verify whether GRB 050509b follows the Amati relation, and so clarify its nature, we simulated a theoretical spectrum (see figure \ref{fig:spectrum}), following scenario 2, to verify a posteriori the consistency of this source with the Amati relation.

We first calculate $E_{iso}$, which, as mentioned above, in our case is not the total energy of the GRB but the total energy of the extended afterglow, $E_{iso}\equiv E_{after}=72\% E_{tot}^{e^\pm} = 4\times 10^{48}$ erg. To calculate $E_{p,i}$, we simulated the $\nu F\nu$ theoretical spectrum integrated over the entire extended afterglow up to $10^4$ s, as described in \citet{2008A&A...487L..37G}. The theoretical $\nu F\nu$ spectrum in the observer frame is shown in figure \ref{fig:spectrum}: it peaks at $E_p \sim 2.3$ keV, which implies that $E_{p,i} = (1+z) E_p \sim 2.8$ keV.

We also checked the position of GRB 050509b in the $E_{p,i}/E_{iso}$ plane considering only the short hard spikelike emission observed by BAT, which is identified with the P-GRB. In this case, only a lower limit to $E_{p,i}$ can be established from the observational data. The $\nu F_\nu$ observed spectrum in the BAT $15$--$150$ keV energy range indeed increase with energy and does not exhibit any peak \citep{2006ApJ...638..354B}. Therefore, a first estimate would lead us to conclude that $E_p > 150$ keV and then that $E_{p,i} > 184$ keV. The corresponding value of the isotropic equivalent energy emitted in the BAT $15$--$150$ keV energy range is $E_{iso,15-150} = (2.7 \pm 1) \times 10^{48}$ ergs \citep{2006ApJ...638..354B}. \citet{2006ApJ...638..354B} conclude that the total isotropic equivalent energy emitted can be $E_{iso} \gtrsim 3E_{iso,15-150}$ if $E_p \gtrsim 1$--$2$ MeV. A more conservative estimate of $E_{p,i}$ and $E_{iso}$ can be found by fitting the observed BAT spectrum with a Band model where $\alpha$ and $\beta$ indices are fixed to typical values ($\alpha = -1$ and $\beta = -2.3$). This leads to the following lower limit to $E_p$ at $90$\% c.l. of $E_p > 55$ keV, which corresponds to $E_{p,i} > 67$ keV. To compute the corresponding total isotropic equivalent energy $E_{iso}$, we must integrate this Band spectrum from $1$ keV to $10000$ keV. Since the exact value of $E_p$ is not known, but we have only a lower limit, the result of this integration, and therefore $E_{iso}$, will depend on $E_p$. We find that $E_{iso}$ can range from $5 \times 10^{48}$ erg, if $E_p$ is equal to its lower limit (i.e. $E_p = 55$ keV), all the way up to $3 \times 10^{49}$ erg, if $E_p$ is as high as the upper limit of the integration (i.e. $E_p = 10000$ keV).

\begin{figure}
\centering
\includegraphics[width=\hsize]{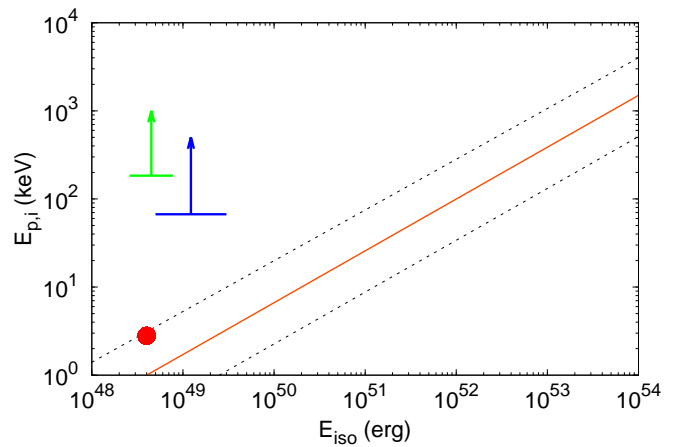}
\caption{GRB 050509b position in the $E_{p,i}/E_{iso}$ plane. The continuous orange lines show the best-fit power law of the $E_{p,i}-E_{iso}$ correlation and the dotted gray ones are the 2$\sigma$ confidence region, as determined by \citet{2009A&A...508..173A}. The green lower limit ($2.6\times 10^{48} < E_{iso} \lesssim 7.8 \times 10^{48}$ erg and $E_{p,i} > 184$ keV, computed following \citealp{2006ApJ...638..354B}) and the blue lower limit ($5 \times 10^{48} < E_{iso} < 3 \times 10^{49}$ erg and $E_{p,i} > 67$ keV, computed following the more conservative approach described in this paper) correspond to the short hard spikelike emission observed by BAT. The red dot ($E_{iso} \sim 4 \times 10^{48}$ erg, $E_{p,i} \sim 2.8$ keV) corresponds to the unobserved peak of the extended afterglow theoretically computed within the fireshell model.}\label{fig:amati}
\end{figure}

In Fig. \ref{fig:amati}, the result of this analysis are shown. When considering both the upper limit following \citet{2006ApJ...638..354B}, namely $E_{p,i} > 184$ keV and $2.6\times 10^{48} < E_{iso} \lesssim 7.8 \times 10^{48}$ erg, and the more conservative upper limit computed above, namely $E_{p,i} > 67$ keV and $5 \times 10^{48} < E_{iso} < 3 \times 10^{49}$ erg, we find that the short hard spikelike emission observed by BAT, which is identified with the P-GRB, does not fulfill the Amati relation. When, instead, we consider the peak of the extended afterglow alone ($E_{p,i} \sim 2.8$ keV, $E_{iso} \sim 4 \times 10^{48}$ erg, see above), as should be done according to the fireshell scenario \citep{2008A&A...487L..37G,2010A&A...521A..80C}, GRB 050509b is fully consistent with the Amati relation.

This result allows us to conclude that, within the theoretical fireshell model, GRB 050509b is consistent with the Amati relation. It implies that this is not a genuine short burst, but instead a long burst disguised as a short one, confirming our hypothesis.

\section{Discussion}\label{sec:discussions}

In a set of papers based on the fireshell model, it has been introduced a new class of GRBs, called ``disguised'' short GRBs \citep[see][and references therein]{2007A&A...474L..13B,2008AIPC..966....7B,2009A&A...498..501C,2010A&A...521A..80C,2009AIPC.1132..199R}. These are canonical long GRBs with an extended afterglow that is energetically predominant with respect to the P-GRB (see Fig. \ref{B}). Their main characteristic is that the emission of the afterglow occurs in an environment characterized by a peculiarly low value of the average CBM density ($n_{CBM}=1.0 \times 10^{-3}$ particles/cm$^3$), which is a value typical of a galactic halo environment. Under this condition, the extended afterglow peak luminosity is much lower than the one expected for a canonical value of the average CBM density inside the galaxy ($n_{CBM}=1.0$ particle/cm$^3$). Consequently, the extended afterglow peak luminosity is ``deflated'' and the energy of the extended afterglow is released on a much longer timescale. The energetic predominance of the afterglow with respect to the P-GRB is quantified by the value of the baryon loading $B$ (see Fig. \ref{B}) and can be verified by integrating over time the luminosity of the extended afterglow. Examples of this class are GRB 970228 \citep{2007A&A...474L..13B}, GRB 060614 \citep{2009A&A...498..501C}, and GRB 071227 \citep{2010A&A...521A..80C}.

We have shown that GRB 050509b cannot be considered a ``genuine'' short GRB and proposed that it be classified as a disguised short GRB. We have tested two alternative scenarios for this GRB, one assuming it is a ``genuine'' short GRB and an alternative one assuming it is a disguised short GRB. We have demonstrated that the only interpretation of the data compatible with the first scenario would lead to an extremely intense P-GRB that is not observed (see Fig. \ref{fig:After}); therefore this scenario should be discarded and GRB 050509b cannot be interpreted as a ``genuine'' short GRB. We have instead obtained a reasonable interpretation within the second scenario for $B = 6.0 \times 10^{-4}$, corresponding to a long GRB, and an average CBM density of $n_{CBM}=1.0 \times 10^{-3}$ particles/cm$^3$, which clearly implies that GRB 050509b is a disguised short GRB (see Fig. \ref{fig:P-GRB}).

GRB 050509b does not have XRT data before $100$ s. The BAT light curve went beneath its threshold at $\sim 40$ ms. All the data about the peak of the extended afterglow is therefore missing, which is the relevant part for the calculation of the energy peak $E_{p,i}$ of the $\nu F\nu$ spectrum for the Amati relation. Despite this lack of data, it has been possible, from our theoretical simulation, to infer both the spectrum of the extended afterglow peak emission and a value of $E_{p,i}$, and to check a posteriori whether GRB 050509b fulfills the Amati relation. This has been done, as already shown in previous papers \citep{2008A&A...487L..37G,2010A&A...521A..80C}, by duly neglecting the contribution of the P-GRB, assuming that the Amati relation is connected only to the extended afterglow emission process. It has been proven indeed (see Fig. \ref{fig:amati}) that GRB 050509b, when the P-GRB contribution in neglected, is in perfect agreement with the Amati relation. This interpretation is also supported by the P-GRB alone being inconsistent with the Amati relation \citep[see also Fig. \ref{fig:amati}, Sec. \ref{amati} and e.g.][and references therein]{2006ApJ...638..354B}.

The understanding reached for this source and others of the same class points also to a difficulty in identifing a ``genuine'' short GRB. A selection effect is at work: a genuine short GRB must have a very weak extended afterglow (see Fig. \ref{B}); consequently, it is very difficult to determine its redshift.

\section{Conclusion}\label{sec:conclusions}

It has been shown that GRB 050509b originates from the gravitational collapse to a black hole of a merging binary system consisting of two degenerate stars according to three different and complementary considerations:
\begin{enumerate}
\item Very stringent upper limits on an associated supernova event have been established \citep[see][]{2005GCN..3401....1C,2005GCN..3521....1B,2005ApJ...630L.117H,2005A&A...439L..15C,2005GCN..3386....1B,2005GCN..3417....1B,2006ApJ...638..354B};
\item The host galaxy has been identified with a luminous, non-star-forming elliptical galaxy \citep{2005GCN..3390....1P,2005Natur.437..851G,2005GCN..3386....1B,2006ApJ...638..354B};
\item The GRB exploded in the halo of the host galaxy \citep{2006ApJ...638..354B}, because the binary system spiraled out before merging.
\end{enumerate}
From an astrophysical point of view, there are three possible cases of merging binary systems that must be considered:
\begin{enumerate}
\item Neutron star / neutron star: unlike the case of GRB 970228 \citep{2007A&A...474L..13B}, the low energetics of GRB 050509b disfavor this hypothesis;
\item Neutron star / white dwarf: this appears to be the most likely case for GRB 050509b, as in GRB 060614 \citep{2009A&A...498..501C} and in GRB 071227 \citep{2010A&A...521A..80C};
\item White dwarf / white dwarf: this case is viable only for two very massive white dwarfs, allowing the critical mass of neutron stars against gravitational collapse to a black hole to be overcome in the merging process; that low massive white dwarf / white dwarf merging binary systems may lead to low energetics events has been widely expressed in the literature \citep[see e.g.][]{1984ApJS...54..335I,1985ASSL..113....1P,2010Natur.463...61P}.
\end{enumerate}
From the point of view of GRB classification, we conclude that:
\begin{enumerate}
\item GRB 050509b is a disguised short GRB occurring in a low CBM density environment ($n_{CBM} < 10^{-3}$ particles/cm$^3$), typical of a galactic halo;
\item The baryon loading of GRB 050509b, and consequently the ratio of the P-GRB to the extended afterglow energetics, is typical of canonical long-duration GRBs;
\item The possible origin of a genuine short GRB from a merging binary system, as often purported in the literature (see e.g. \citealp{2006RPPh...69.2259M} but also \citealp{2009ARA&A..47..567G}), still remains an open issue both from an observational and a theoretical point of view; in theory, this will crucially depend on the amount of baryonic matter left over in the process of gravitational collapse originating the fireshell baryon loading, which must be $B \lesssim 10^{-5}$.
\end{enumerate}
From all the above considerations, it also follows that a binary system merging in a higher density region (i.e. $n_{CBM} \sim 1$ particles/cm$^3$) would give rise to a canonical long-duration GRB without an associated supernova \citep[see also][]{2007A&A...474L..13B,2009A&A...498..501C}.

\acknowledgements

We thank Dr. Cristiano Guidorzi and Dr. Raffaella Margutti for their support in the data analysis. We thank as well an anonymous referee for her/his important suggestions which improved the presentation of the manuscript. The support of the IRAP PhD program is acknowledged by GDB and LI.

\bibliographystyle{aa}
\bibliography{GRBs,Neo-GRBs,Paleo-GRBs}

\end{document}